# To Switch or Not to Switch – a Machine Learning Approach for Ferroelectricity

Sabine M. Neumayer,[a] Stephen Jesse,[a] Gabriel Velarde,[b] Andrei L. Kholkin,[c] Ivan Kravchenko,[a] Lane W. Martin,[b] Nina Balke[a] and Peter Maksymovych[a*]

[a] Center for Nanophase Materials Sciences, Oak Ridge National Laboratory, Oak Ridge, TN 37831, USA

[b] Department of Materials Science and Engineering, University of California, Berkeley, CA 94720, USA & Materials Sciences Division, Lawrence Berkeley National Laboratory, Berkeley, CA 94720, USA

[c] Department of Physics & CICECO – Aveiro Institute of Materials, University of Aveiro, Aveiro, Portugal & Scool of Natural Sciences and Mathematics, Ural Federal University, Ekaterinburg, Russia

*maksymovychp@ornl.gov

*With the advent of increasingly elaborate experimental techniques in physics, chemistry and materials sciences, measured data are becoming bigger and more complex. The observables are typically a function of several stimuli resulting in multidimensional data sets spanning a range of experimental parameters. As an example, a common approach to study ferroelectric switching is to observe effects of applied electric field, but switching can also be enacted by pressure and is influenced by strain fields, material composition, temperature, time, etc. Moreover, the parameters are usually interdependent, so that their decoupling toward univariate measurements or analysis may not be straightforward. On the other hand, both explicit and hidden parameters provide an opportunity to gain deeper insight into the measured properties, provided there exists a well-defined path to capture and analyze such data. Here, we introduce a new, two-dimensional approach to represent hysteretic response of a material system to applied electric field. Utilizing ferroelectric polarization as a model hysteretic property, we demonstrate how explicit consideration of electromechanical response to two rather than one control voltages enables significantly more transparent and robust interpretation of observed hysteresis, such as differentiating between charge trapping and ferroelectricity. Furthermore, we demonstrate how the new data representation readily fits into a variety of machine-learning methodologies, from unsupervised classification of the origins of hysteretic response via linear clustering algorithms to neural-network-based inference of the sample temperature based on the specific morphology of hysteresis.*

## Introduction

Nano- and mesoscale electromechanical behavior underpins the performance of sensors, actuators, energy harvesters, ferroelectric field effect transistors and electrocaloric devices.[1-9] Interesting phenomena arise from local defect chemistry, chemical domains and interfaces such as domain walls, grain boundaries and the surface itself, necessitating local probing techniques to study functional material response.[10-20] Therefore, piezoresponse force microscopy (PFM) and related techniques that probe piezoelectric properties on the nm to μm scale have become increasingly popular.[21-26] In these techniques, AC and DC voltages are applied to a conductive tip in contact with the sample surface and the electromechanical response of the surface to the voltage stimuli is detected. The periodic cantilever deflection $D_{ac}$ in response to the applied voltage provides information on the strength of electromechanical interaction and the direction of the ferroelectric polarization.

Typically, electromechanical response is studied as a univariate function of DC voltage, producing a characteristic hysteresis loop which is reminiscent of macroscopic polarization-switching measurements. Hysteresis exhibits variability, however, with the number of voltage cycles, time and temperature. In order to analyze and extract meaning from these data sets, machine learning has become increasingly important.[27-34] Dimensionality reduction without loss of important information, de-noising, clustering and identifying characteristic features in data sets have been achieved using supervised and unsupervised machine-learning algorithms.

Despite the growing number of successful statistical analyses, the previous efforts to probe and analyze hysteresis attempt to separate interdependency of control parameters, typically by varying one parameter while keeping the rest of the conditions constant (for example measuring piezoresponse at a fixed voltage along the switching cycle – so-called switching spectroscopy PFM[35]). To what extent such decoupling of control variables can be systematically achieved is up to debate. But more importantly, it is natural to ask whether there exist more encompassing approaches, where coupling of control parameters can be explicitly investigated.

For a model setting, we focus on hysteresis measured with the technique of contact Kelvin probe force microscopy[36, 37] (cKPFM). cKPFM is conceptually analogous to commonly used open-loop Kelvin probe force microscopy (KPFM), albeit with the tip in mechanical contact with the measured surface. As is the case with KPFM, cKPFM detects net changes of electrostatic forces, this time acting on the leads of the capacitor, that sandwiches the studied dielectric between the scanning probe and the bottom electrode. Although less sensitive than KPFM owing to much larger contact stiffness in comparison to the unclamped cantilever, the demonstrated utility of cKPFM is to tease out electrostatic contributions to the net force acting on the capacitor, which can be complementary or competing to other forces – for example piezoelectric deformation of the dielectric. Therefore, cKPFM found application in distinguishing between piezoelectric and electret response of dielectrics to applied voltage,[38] particularly relevant for new and/or nanoscale materials, where piezoelectric response can be small and electromechanical properties unknown.

Here we introduce a two-dimensional (2D) representation of the field-induced hysteresis measured by cKPFM, which explicitly considers two-parameter dependence of the net measured signal on the "write" and "read" bias signals. Remarkably, just the transition to 2D dependencies and representation vastly simplifies qualitative interpretation of the measured signal, clearly separating, for example, ferroelectric and electret behaviors. Subsequently, we demonstrate how this 2D approach holistically facilitates application and interpretation of a variety of machine-learning algorithms, including artificial neural networks (ANN), that extend the applicability of this methodology to detect incomplete switching and ferroelectric-relaxor phase transitions with minimum human input. Even inference of the temperature of the sample from its hysteretic response appears to be possible with fairly simple neural networks. We believe that this data representation technique will help advance the experimental methodologies of hysteretic spectroscopy and spectro-microscopy to characterize memory functions and hysteretic materials in general and will also help bridge the theory and experiment in a more statistically robust setting.

**Results and discussion**

Hysteresis loops measured with cKPFM represent a basic example of dependency of the response on a combination of two electrical signals – one required to "write" the hysteretic state of the system, and the other one – to "read" the state. And, somewhat ironically, this two-parameter dependence presents one of the primary difficulties with observational interpretation of the cKPFM data. Indeed, a successful interpretation requires considering not only the opening of the hysteresis loop, but also the evolution of the loop with varying read-bias. Without rigorous statistical metrics, such an interpretation is quite challenging, especially in cases where electromechanical and electrostatic signal contributions are of comparable strength. This is particularly true for intermediate cases where subtle as well as dramatic features may arise in the hysteresis loop.

To demonstrate our approach, we utilize cKPFM response for different ratios of $V_{read}$ and $V_{write}$ on a model ferroelectric lead zirconium titanate (PbZr$_{0.2}$Ti$_{0.8}$O$_3$, PZT) film before discussing typical cKPFM response on a non-piezoelectric amorphous hafnium dioxide (aHfO$_2$) sample, where electrostatic effects can lead to observation of hysteresis loops in standard switching spectroscopy PFM.[39] In a next step, we analyze spatially varying cKPFM response measured on macroscopically pre-poled lithium niobate (LN) that is subject to strong electrostatic interactions typical for that ferroelectric material.[10] To further corroborate the applicability of our analysis approach we process cKPFM response measured across the ferroelectric – relaxor phase transition on multiple grains of lanthanum zirconate titanate (PLZT), which even in the ferroelectric state exhibits peculiarities in hysteresis loop, as commonly observed for relaxors.[31]

Figure 1(a) schematically depicts the DC voltage waveform used in cKPFM, which consists of triangular write pulses $V_{write}$ to initiate ferroelectric switching and a probe voltage $V_{read}$ that is applied between the write pulses and stepwise changed with every write cycle. Therefore, the measured response is a function of read and write voltage and spans a 2D parameter space for each probed location. The cKPFM data (Figure 1) were measured on PZT, one of the most common ferroelectrics.

In the previous representation of the cKPFM data (Figure 1(b)), response $D_{ac}$ is depicted versus $V_{read}$ and data acquired during each $V_{write}$ step is overlaid on the same plot, with different line colors indicating the preceding write step.[37, 39] Interpretation of these diagrams, however, can be challenging. cKPFM diagrams typically show 40-100 lines in one graph, which often overlap and can therefore be hard to interpret. Moreover, it can be difficult to exactly distinguish between color nuances that correspond to the preceding $V_{write}$ step. The classification as ferroelectric switching or electrostatically driven artefacts, which is often the main reason to apply this technique, has been only vaguely and qualitatively defined as the presence/absence of formation of "two bands" and the cKPFM curve shapes resembling a hysteresis loop, as opposed to electrostatic artefacts which would appear as one band of straight lines.[10, 39] Such an assignment is highly dependent on the probing versus writing voltage ratios, as well as the vagueness of the definition of a "good" hysteresis loop.

Our proposed alternative is a 2D representation of the cKPFM experiment, which (1) unfolds the individual hysteresis loops so that they are a function of time or voltage step, rather than applied bias; and (2) stacks the progression of the unfolded loops into a matrix, where each row corresponds to a certain $V_{read}$ and each column to a certain $V_{write}$ step. The unfolding and stacking procedure is shown in Figure 1 (c-d). The strength of the response at each $V_{read}$/$V_{write}$ parameter pair value is represented by the color scale. The magnitude of $V_{write}$ is indicated by the white, dashed triangular line for reference purposes. Slicing the cKPFM map along the columns corresponds to $D_{ac}$ as a function of $V_{read}$, *i.e.*, the traditional cKPFM representation. A single slice at $V_{read} = 0$ (followed by conversion of the x-axis to $V_{write}$) corresponds to the often used "remnant" switching spectroscopy hysteresis loop. The diagonal slice at $V_{read} = V_{write}$ produces the "in-field" hysteresis loop. Of course, many other forms of 1D hysteresis can also be created, via cuts in arbitrary directions in $V_{read}$ -$V_{write}$ space.

Therefore, right away, the cKPFM map generalizes the measurement of the hysteretic response of the dielectric to applied stimuli, encompassing, in principle, the response to all possible combinations of read-write waveforms.

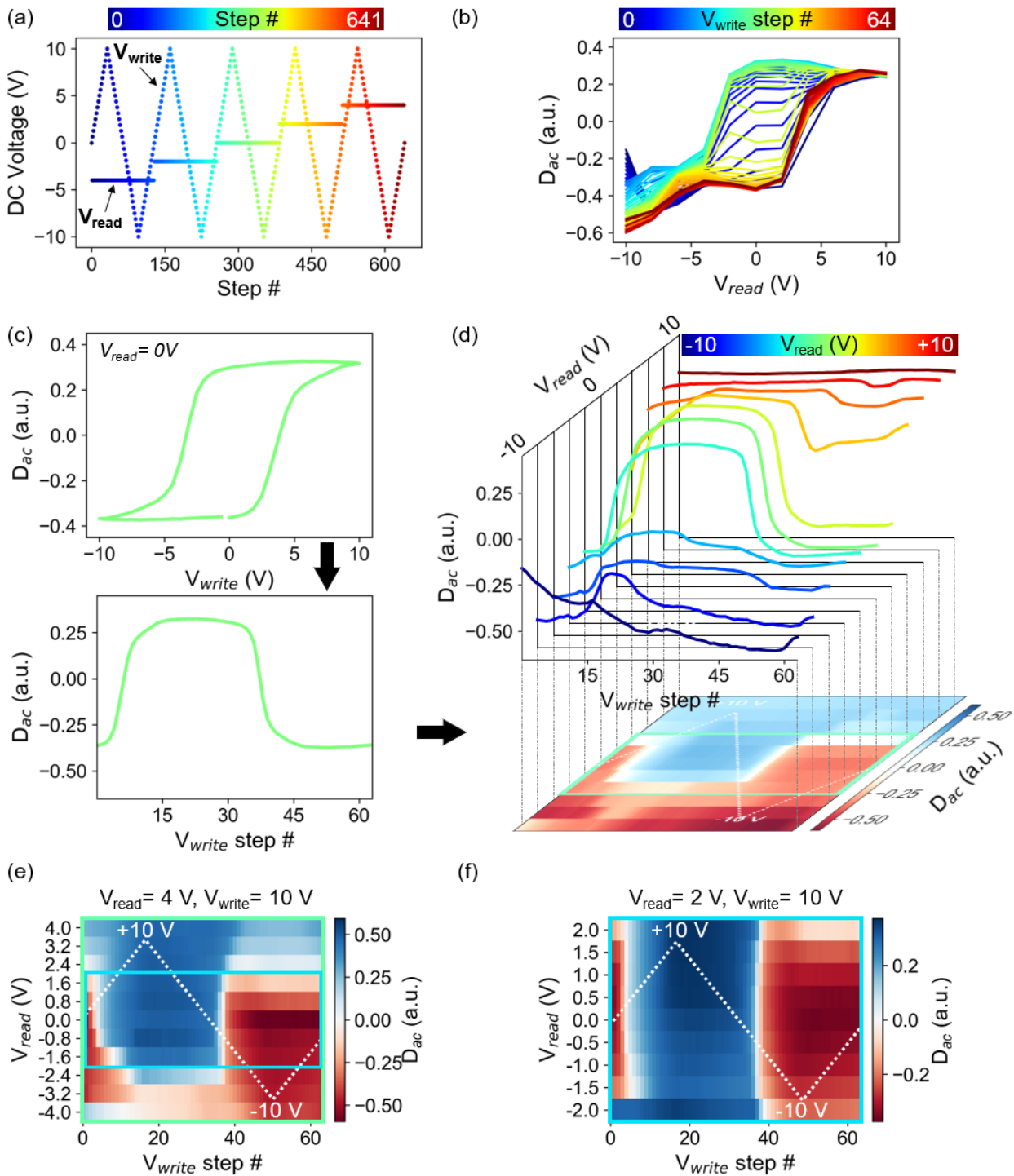

**Figure 1:** cKPFM on ferroelectric PZT. (a) Schematic of sequence of DC voltage pulses applied during cKPFM. $V_{write}$ pulses are increasing and decreasing in a triangular envelope, whereas the $V_{read}$ is applied between write pulses and sequentially increased with each cycle. (b) Traditional cKPFM diagram where response $D_{ac}$ is plotted as a function of $V_{read}$ with $V_{write}$ steps color coded. (c) Hysteresis loop extracted from $V_{read}$ = 0 as a function of $V_{write}$ (top) and unfolded as a function of $V_{write}$ step #. (d) Unfolded loops for all $V_{read}$ stacked along a third dimension. These loops are projected on a 2d map where $D_{ac}$ is represented by the color bar, rows correspond to response during a certain $V_{read}$ and columns correspond to $V_{write}$ steps. (e) cKPFM with decreased $V_{read}$ window as indicated in the green box in panel (d), (f) cKPFM map with a further decreased the probing window corresponding to the blue box in panel (e). The white dashed line in cKPFM maps in panels (d-f) indicate $V_{write}$.

For the specific case of PZT, the cKPFM map shows sharp horizontal and vertical edges where the measured $D_{ac}$ abruptly changes, indicating ferroelectric switching events. Obviously, because $V_{write}$ and $V_{read}$ both apply electric field across the dielectric, the response is not fully independent as a function of these parameters. For example, if $V_{read}$ exceeds the coercive voltage, polarization will necessarily switch into the preferred orientation. This is shown in Figures 1(d), (e) and (f), where $V_{read}$ was changed from values between ±4 V (Figure 1(e)) and ±2 V (Figure 1(f)), while $V_{write}$ was kept the same triangular wave between ±10 V. The coercive voltages are approximately 4 V for the film (Figure 1(c)). Thus, switching is induced by both $V_{read}$ and $V_{write}$ in Figure 1(e) whereas in Figure 1(f), only $V_{write}$ is ramped above the coercive voltage. This is not, however, a limiting factor for statistical analysis, since a properly trained classifier can easily incorporate the additional information on when switching occurs during the $V_{write}$ and $V_{read}$ steps. One point of potential inquiry is the time-dependence of the cKPFM maps. We will not address it here, but naturally there is always some level of time-dependence for hysteretic processes, and under the right approach this will enhance the level of understanding even more. For now, we will proceed with the assumption that cKPFM maps are not strongly time-dependent, representing the case of measurement much slower than the characteristic response time of the material.

The immediate utility of cKPFM maps is that they represent spectral information as images. Therefore, the data can be more intuitive to interpret for the human mind and is more suitable for image-based deep-learning algorithms that are becoming increasingly important. To illustrate some of these advantages, we begin with the simplest task of data interpolation.

For experimental reasons, such as acquisition time and wear of the conductive tip coating, response maps can only be acquired for a limited amount of sampling points in voltage space. 2D interpolation, however, provides higher-resolution maps from which gradients can be obtained to further highlight response function characteristics. Interpolation can be calculated with a wide variety of algorithms (*e.g.*, Gaussian progress regression, linear, cubic, spline), many of which are conveniently implemented in open-source packages such as scikit-image, scipy.interpolate, etc.[40, 41]

Interpolation of cKPFM maps for ferroelectric PZT and non-ferroelectric aHfO$_2$ are shown in Figure 2. A simple check of the quality of the interpolation, other than the 2D image itself, is seen in the extracted 1D remnant hysteresis loop (supplementary information, Figure S1), revealing arguably excellent quality of interpolated values. With the interpolated values, one can extract gradients of the dielectric response to $V_{read}$ and $V_{write}$. The gradients present a transparent and simple approach to differentiate characteristic behaviors, *e.g.*, to compare piezoresponse of ferroelectric to electrostatic forces of a leaky dielectric.

Figure 2 shows experimental data, interpolated maps and the gradients calculated from the interpolated data for ferroelectric PZT using $V_{read}$ amplitudes of 2 V and $V_{write}$ amplitudes of 10 V (Figure 2(a)) and non-ferroelectric aHfO$_2$ (Figure 2(b)). The corresponding gradients reveal sharp edges in the x-gradient for ferroelectric switching and little contrast in the y-gradient, as expected, based on abrupt switching of the sign of the piezoelectric response at specific voltages on one hand, and weak dependence of piezoelectric response with applied voltage on the other. A composite figure showing cKPFM maps of experimental data, interpolated data, x- and y-gradients for PZT upon applying different $V_{read}$ and $V_{write}$ ratios is shown in the supplementary information, Figure S2. An interesting case is the hysteresis of aHfO$_2$, which previously was assigned to transient charging of the probed volume by rechargeable traps.[36] The aHfO$_2$ cKPFM map is sharply contrasting that of a ferroelectric. Figure 2(b) shows that there is no abrupt change of contrast in the x-direction (of varying $V_{write}$), but rather a gradual wave-like modulation of the response along the y-direction (of varying $V_{read}$). The corresponding gradients in x-direction are very smooth (unlike the sharp transitions of the ferroelectric case). Yet, there is also a notable feature in the y-gradient, corresponding to the position of the contact potential minimum, that shifts with applied $V_{write}$. Applying higher read and write voltages to increase charge injection does not significantly change the overall contrast of experimental and interpolated data maps and x-gradients still do not exhibit sharp vertical contrast but rather exhibit V-shaped tilted lines of similar values as the background (see Figure S3 in the supplementary information). Therefore, cKPFM maps in the absence of polarization switching as in the presented aHfO$_2$ data are very clearly distinguishable from the case of ferroelectric switching, and reveal useful features, such as the modulation of the contact potential difference.

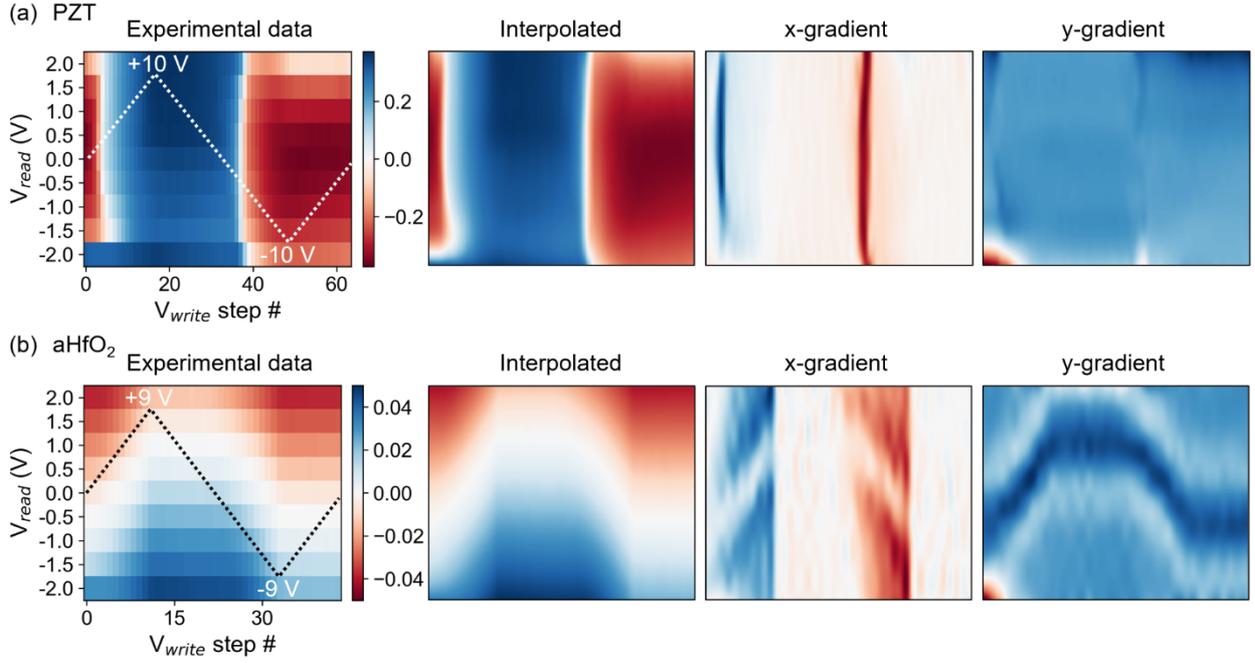

**Figure 2:** Maps of experimental cKPFM data, interpolated data and the gradients in x- and y-direction calculated from interpolated maps. (a) For PZT where ferroelectric switching events are characterized by the sharp contrast transition corresponding to thin lines in the x-gradient map and (b) for non-ferroelectric aHfO$_2$ where D$_{ac}$ changes gradually dependent on Vread and V$_{write}$, resulting in broad bands in the x-gradient map.

Due to the distinctive features of non-switching versus switching cKPFM data, statistical clustering techniques can be utilized to identify functional behavior in large data sets dependent on material, spatial region within a sample, temperature or time. In general, clustering algorithms find similarities in data and use those similarities to group data. Two examples for suitable methods to break down the different types of detected response are hierarchical agglomerative clustering (HAC) and density based spatial clustering (DBSC).[42] These algorithms can either be applied directly to the measured response or after de-noising and dimensionality reduction, *e.g.*, through principal component analysis (PCA).

HAC initially assigns a cluster to each cKPFM map in a given data set. In a next step, the most similar clusters where the response is closest in the feature space are paired. These grouped clusters are then clustered again dependent on their similarities. Clustering into a decreasing number of clusters containing an increasing amount of cKPFM maps is continued iteratively until all the data is grouped into two main clusters. We apply HAC directly to a data set consisting of cKPFM data for three types of behavior: (i) electret response measured on aHfO$_2$, (ii) ferroelectric switching measured on PZT and (iii) response from ferroelectric PZT upon applying DC voltages below the coercive voltage, which does not initiate polarization switching. The total number of data sets was 3000, comprising 400 measured on PZT upon switching with different ratios V$_{read}$/ V$_{write}$ ratios, 100 data sets for sub-coercive voltages on PZT and 2500 measured on aHfO$_2$ representing the electret response. The results of HAC are depicted in Figure 3. The dendrogram in Figure 3(a) shows the relationship between the identified clusters with the vertical lines indicating similarities between the grouped data. The shorter the vertical line, the more similar is the response. The numbers in parentheses on the x-axis correspond to the number of cKPFM maps within the clusters associated with the vertical lines. The two most distinct clusters are indicated in blue in the dendrogram and represent the cKPFM data in Figure 3(b) for point 1 and 2 at a numeric distance of 150. Clearly, cluster #1 exhibits non-ferroelectric behavior similar to the map for aHfO$_2$ depicted in Figure 3(b). Contrary, cluster #2 corresponds to ferroelectric switching associated with the sharp contrast as discussed previously. Thus, unsupervised clustering immediately answers the proverbial "To switch or not to switch" question, which continues to dominate the studies of nanoscale ferroelectrics, 2D materials and other emerging members of the ferroelectric family, where the signals are typically quite weak.[38]

If we further look into the next level of clustering, however, both clusters #1 and #2 separate into methodologically meaningful and useful categories. Cluster #1 correctly subdivides cKPFM data where no polarization switching occurs either due to sub-coercive voltages (cluster #5) or a non-ferroelectric sample where contrast is governed by electrostatic interactions (cluster #6), as shown in Figure 3(b). Indeed, although the material is itself ferroelectric, this fact cannot be inferred with sub-coercive voltage spectroscopy, which generates no switching by electric field (see maps in Figure S2, first row). At the same time, cluster #2, partitions the data depending on the extent by which $V_{read}$ exceeds the coercive voltage (cluster #3 and #4), which is a valuable methodological distinction, as discussed earlier. Due to the hierarchical approach, clusters #3 to #6 can be subdivided into more clusters that show similar behavior to each of the parent clusters.

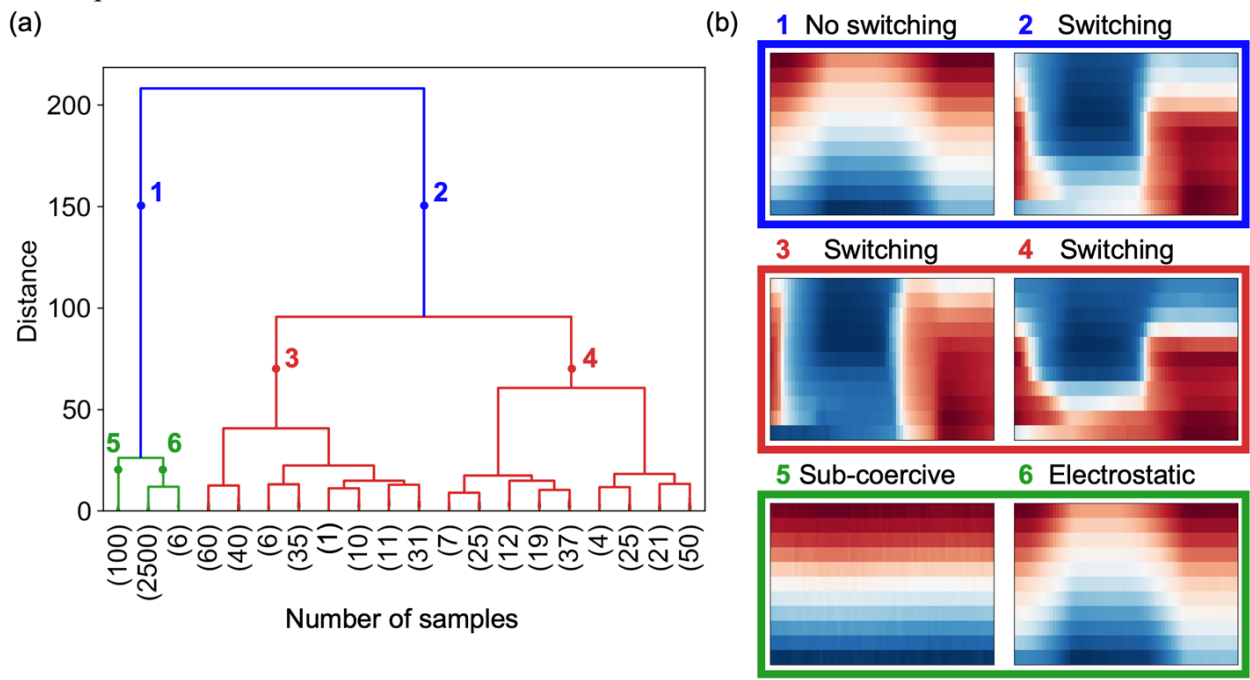

**Figure 3:** Hierarchical agglomerative clustering. (a) Dendrogram indicating distribution of clusters and their similarities (indicated by the length of vertical lines), (b) cKPFM maps of mean response for clusters indicated by color and numbers in panel (a). Cluster #1 identifies characteristics where no polarization switching occurs either because a $V_{write}$ below the coercive voltage is applied to a ferroelectric sample (cluster #5) or the response is purely electrostatic (cluster #6). Cluster #2 groups data of switching events occurring at different $V_{write}$ steps (cluster #3 and #4).

Apart from grouping data of different materials, clustering is also able to identify local variability of hysteretic response within a single sample. To this end, we combine PCA and clustering to categorize cKPFM data measured on a spatial grid of 40x40 pixels on LN. The LN sample had been macroscopically pre-poled and dynamic response on this material is typically subject to strong electrostatic interactions even in the presence of polarization switching.[10]

In a first step, PCA projects cKPFM maps into a lower dimensional feature space.[28, 42] This way, component #0 contains the most information on variance in the data, followed by component #1 and so on.

PCA scores are a projection of the data points on the found eigenvectors. Figure 4(a) shows scores for PCA components #0 to #3 represented as 2D cKPFM maps as discussed previously. The scree plot (Figure 4(b)) shows that the variance within the data is successfully captured by about 3 components, with PCA components #0 through #2 accounting for almost 100% of total variance.

In a next step, DBSC is applied to the first three PCA components. DBSC algorithms group data based on the density of data points. Unlike k-means, where the user defined number of clusters dominates the clustering result, the most important parameters in DBSC are the maximum distance between two data points to be considered in the same neighborhood and the minimum number of datapoints within a neighborhood. DBSC on the first three PCA components identified 3 clusters (shown in blue, red and green) in Figure 4(c). Reverting back to the original grid of points across the LN surface, we observe that the clustering identifies distinct hexagonal areas (red) on the surface, while the small green cluster is primarily at the boundary of this area. This observation matches the history of the sample, which was oriented in +z direction and macroscopically pre-poled in hexagonal areas with -z polarization orientation.[10] The mean response for each cluster is shown in Figure 4(e). Clearly, the difference between the clusters corresponds to switching (Cluster #0) and non-switching (Cluster #1 and #2) behaviors within the applied voltage range. Note that while no polarization switching occurs for these clusters, the similarity to cKPFM maps acquired on $aHfO_2$ indicates charge injection and strong electrostatic contributions to the measured signal. The not assigned data points appear to be a mixture between switching and non-switching characteristics and correspond to data points at the decision boundaries in Figure 4(c). Apart from HAC and DBSC, other algorithms like k-means (*e.g.*, Li *et al.*[32], Neumayer *et al.*[31]) can be used to group response to find trends, *e.g.*, dependent on material, location, temperature, etc.

Successful differentiation of switching properties via machine learning of cKPFM maps motivated us to apply non-linear clustering methods, such as neural networks, that could potentially reveal even more details of dielectric behavior. Moreover, while previously discussed clustering algorithms implicitly assume linear separability between members of different clusters in Euclidian space, multilayer perceptron ANNs are able to separate data where this requirement is not fulfilled.[43]

First, we trained an autoencoder network to reveal how effective simple ANN structures are at capturing and reproducing cKPFM maps. Subsequently, we clustered the maps into characteristic types of behavior, using the autoencoder's latent space as a low-dimensional representation of the cKPFM dataset. We utilized temperature-dependent cKPFM data across the ferroelectric-relaxor phase transition of PLZT.[31] The electromechanical response of relaxor ferroelectrics can be particularly challenging to analyze due to their peculiarities in hysteresis loops, even in the ferroelectric state.[44] The proximity of the phase transition point, however, allows us to see how effective machine learning is in capturing the relevant changes of the hysteretic response.

The experimentally acquired cKPFM maps are depicted in Figure 5(a), with the temperature as a third dimension. Figure 5(b) shows the topology of the autoencoder network ANN. We intentionally minimized the complexity of the network, which essentially consists of two sequential multiperceptrons (layers 2 and 4), separated by non-linear activation layers. Yet remarkably, such a network can efficiently capture a high level of detail in the cKPFM map, as shown in Figure 5(c), where we compare a random selection of measured cKPFM maps extracted at different locations and temperatures to their reconstructions by the trained autoencoder.

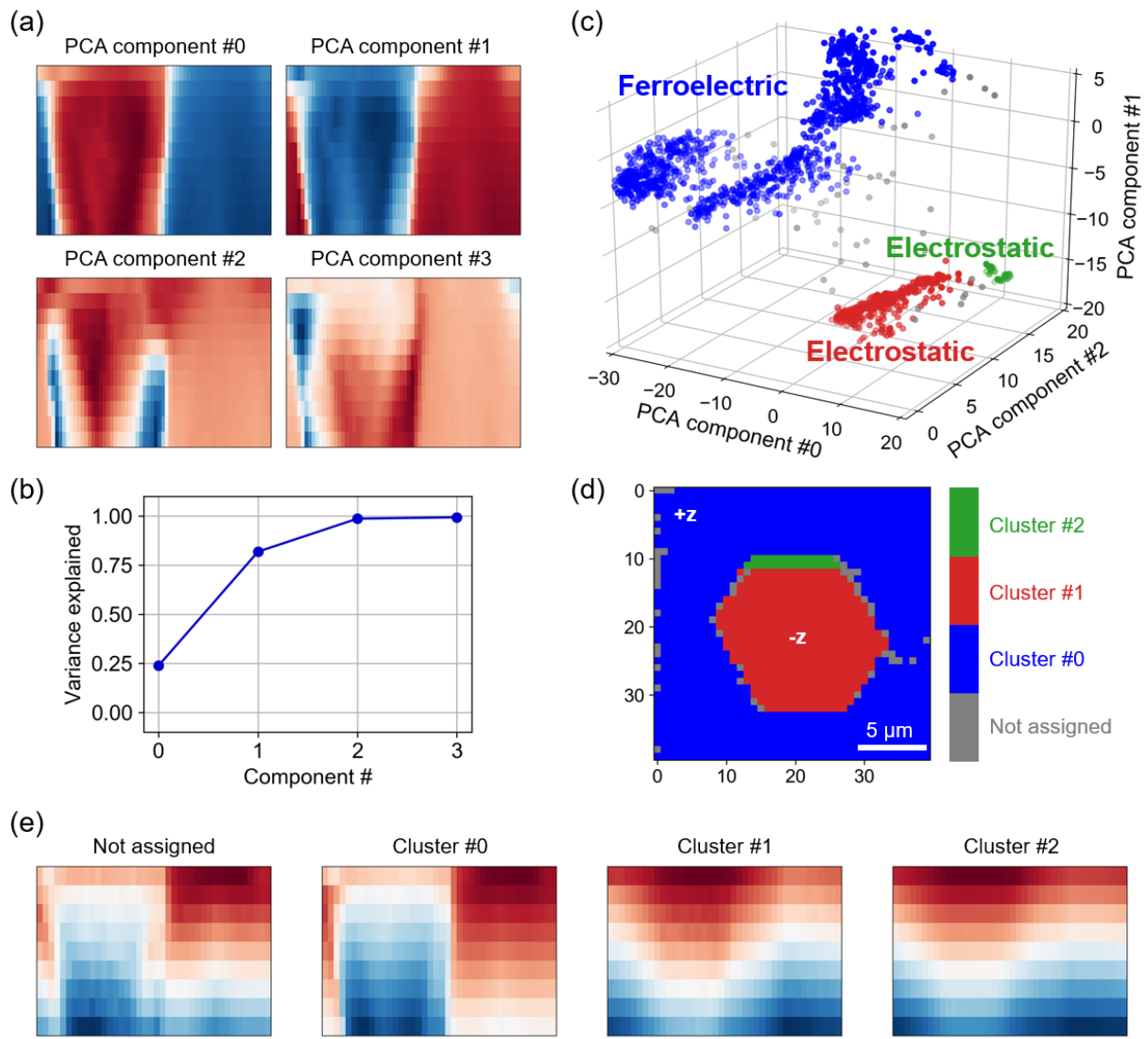

**Figure 4**: Dimensionality reduction and density based spatial clustering of cKPFM measured on LN. (a) Score maps of PCA components and (b) scree plot indicating the data explained with each PCA component. (c) 3d representation of PCA components #0 to #2, color indicates DBSC clusters. (d) Spatial map of DBSC cluster assignment, (e) maps of mean response for each cluster. Cluster #0 (blue) corresponds to ferroelectric switching, cluster #1 and #2 are characteristic for electrostatic response.

Subsequently, we truncated the autoencoder to layer 5, so as to project the measured cKPFM maps onto the encoder's latent 4D space – a common practice for such techniques.

Equally noteworthy is that ANNs can then be used to infer certain experimental conditions from the cKPFM map. To this end, we trained a linear network shown in Figure S4(a) to predict the temperature of the sample from the cKPFM map, using one subset of the temperature-dependent data-set for training of the cKPFM - temperature relationship, and another subset for validation.

.

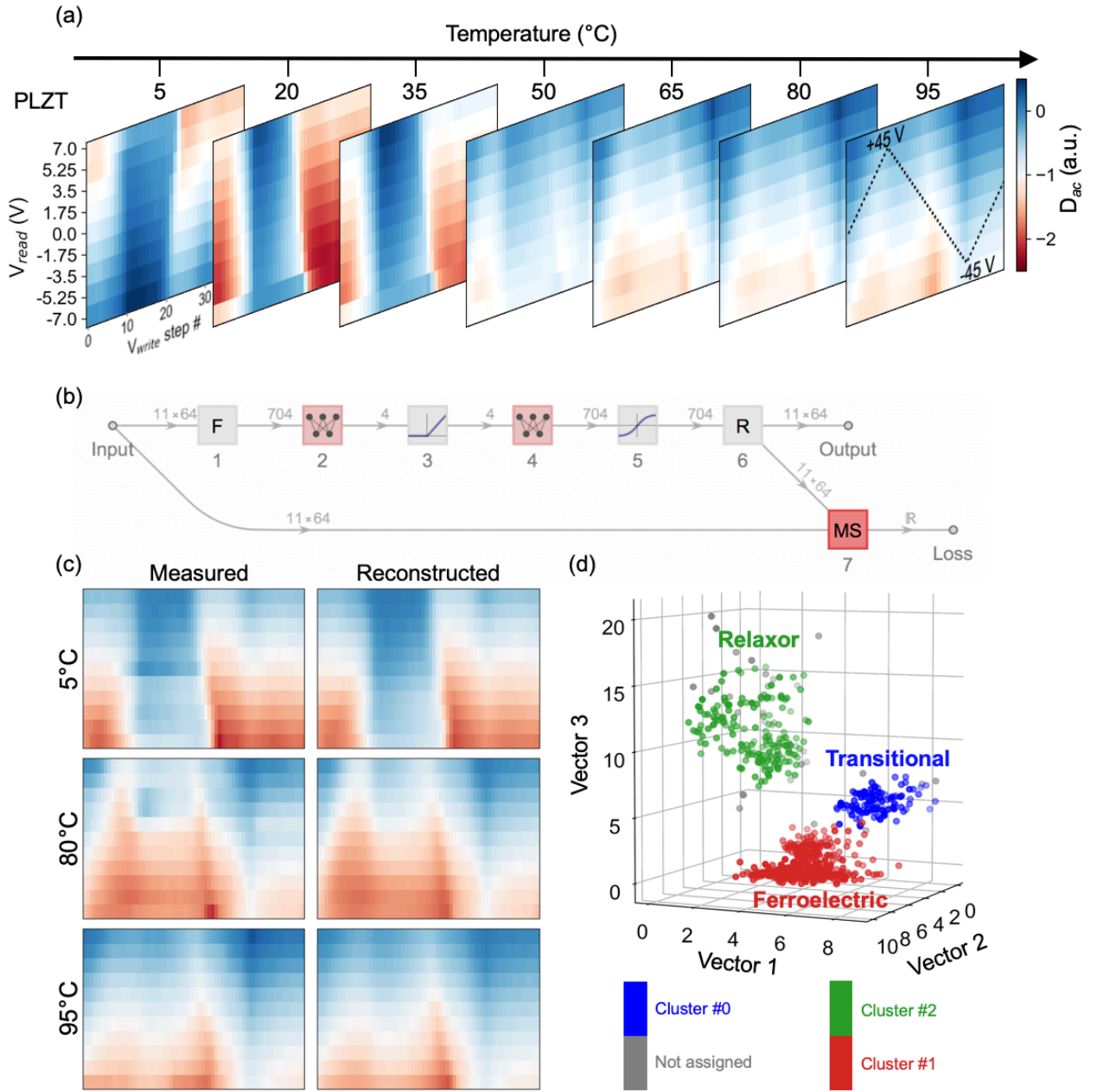

**Figure 5:** Training ANN autoencoder to represent cKPFM data. (a) Experimental cKPFM data on PLZT measured at 7 different temperatures across the ferroelectric-relaxor phase transition. (b) The graph of autoencoder network, starting with flattening of the input array (1), followed by mapping into 4 layer perceptron (2), rectifying activation (3), mapping into 4 layer perceptron (4), sigmoid activation (5) and reshaping (6). "MS" corresponds to mean-squared loss layer. The dimensionality of tensors is shown above the graph edges. (c) Randomly chosen measured and cKPFM maps corresponding to cKPFM data from single pixels at the indicated temperature, (d) DBSC on resulting encoded vectors. Clustering was performed in 4d latent space, for representation purposes only vector 1 to 3 are plotted in the 3d graph.

As seen in Figure S4(b), the network can not only predict the data for the low (5 °C) and high (95 °C) temperatures with a high degree of certainty, but it is also effective in partitioning the whole data set into 5°C increments. The least accurate predictions (with an accuracy of ~ 70%) correspond to temperatures of ~80 °C. But even in this case, the prediction error concentrates within ± 5 °C proximity of the given temperature.

Given the simplicity of the networks employed, we can readily envision a rich spectrum of applications for this type of machine learning, enabled by increasing complexity as well as flexibility of the network functions – for example with the user of convolutional variational autoencoders, as well as a platform for matching models of the electromechanical response to observations

**Conclusions**

In conclusion, we introduced a 2D representation of electromechanical response measured by scanning probe microscopy on ferroelectric and electret samples and demonstrated the ability of machine learning algorithms to distinguish between functional material characteristics. The two-dimensional maps are much easier to interpret by qualitative inspection, than the corresponding one-dimensional representations of dielectric and hysteretic response, and they provide a fertile opportunity for machine-learning techniques to capture and infer material properties. In particular, we provide now statistically robust and automated differentiation between electrostatic charging and ferroelectric switching, which is particularly important for the emerging fields of nanoscale ferroelectrics and energy efficient electronics.     Moreover, even simple neural networks trained on this representation can detect phase transitions across distinct dielectric properties and even infer experimental parameters. While discussed here for cKPFM data, the new representation and its analysis can be extended for other dielectric and electromechanical spectroscopy measurements, such as first-order reversal curves or relaxation data. Moreover, the representation is completely general, and can be easily applied to any dynamic and hysteretic response, with likely effectiveness in systematic analysis of magnetic and resistive hysteresis, as well as higher dimensionality of the measurement, for example to capture relaxation behavior in time as a function of temperature and applied voltages. Overall, we believe that higher dimensional data representation coupled with machine learning will provide numerous advances in microscopy and spectroscopy, particularly in the areas of noisy and complex response as well as theory-experiment matching.

**Methods**

The 150 nm $PbZr_{0.2}Ti_{0.8}O_3$ thin-film was grown on 25 nm of $SrRuO_3$ bottom electrode and $DyScO_3$ (110) single- crystal substrate via pulsed laser deposition. With use of a Kr-F excimer laser (Coherent LPX-300), the $SrRuO_3$ bottom electrode was first deposited at a heater temperature of 640°C in a dynamic oxygen-partial pressure of 100 mTorr with a laser repetition rate and fluence of 14 Hz and 1.23 J cm$^{-2}$, respectively. Next, the chamber was adjusted to a heater temperature of 630°C and dynamic oxygen-partial pressure of 200 mTorr before depositing 150 nm $PbZr_{0.2}Ti_{0.8}O_3$ at a laser repetition rate and fluence of 2 Hz and 1.43 J cm$^{-2}$, respectively. Lastly, the ferroelectric heterostructure was cooled to 25°C at 5°C min$^{-1}$ under a static oxygen pressure of ~700 Torr.
See references for details on the $aHfO_2$,[39] LN[10] and PLZT[31] samples.

cKPFM was measured using Nanosensor PPP-EFM ($aHfO_2$) tips or Budget sensors (all other samples) with a nominal force constant around 3 N/m on commercial atomic force microscopes ($aHfO_2$: Bruker Icon, lithium niobate: Asylum Research MFP-3D, PLZT: Asylum Research,

Cypher). Custom LabView codes and National Instruments data acquisition hardware was used to acquire cKPFM data.

All analysis was performed in Python[45] and Mathematica (ANN).

**Conflicts of interest**

The authors declare no competing interests.

**Acknowledgements**

Experiments on PZT, data analysis and manuscript preparation were supported by the Division of Materials Science and Engineering, Basic Energy Sciences, US Department of Energy (S.N., N.B., P.M.). Experiments were conducted at and supported by (S.J.) the Center for Nanophase Materials Sciences, which is a DOE Office of Science User Facility. S.M.N. would like to thank Katia Gallo and Mohammad Amin Baghban for providing the lithium niobate sample. G.V. acknowledges support from the National Science Foundation under grant DMR-1708615. L.W.M. acknowledges support from the Army Research Office under grant W911NF-14-1-0104. Part of this work was developed within the scope of the project CICECO-Aveiro Institute of Materials, refs. UIDB/50011/2020 & UIDP/50011/2020, financed by national funds through the FCT/MEC.

**SUPPLEMENTARY INFORMATION**

**To Switch or Not to Switch – a Machine Learning Approach for Ferroelectricity**

Sabine M. Neumayer[1], Stephen Jesse[1], Gabriel Velarde[2], Andrei L. Kholkin[3], Ivan Kravchenko[1],

Lane W. Martin[2], Nina Balke[1], Peter Maksymovych[1*]

[1] Center for Nanophase Materials Sciences, Oak Ridge National Laboratory, Oak Ridge, TN 37831, USA
[2] Department of Materials Science and Engineering, University of California, Berkeley, Berkeley, CA 94720, USA & Materials Sciences Division, Lawrence Berkeley National Laboratory, Berkeley, CA 94720, USA
[3] Department of Physics & CICECO -Aveiro Institute of Materials, University of Aveiro, Aveiro, Portugal & School of Natural Sciences and Mathematics, Ural Federal University, Ekaterinburg, Russia

*maksymovychp@ornl.gov

**Section I**

Experimental cKPFM data and interpolation extracted from cKPFM maps shown in Figure 3 for (a) PZT and (b) aHfO$_2$ between data points. Interpolation was performed using the scikit-image transform function.[40]

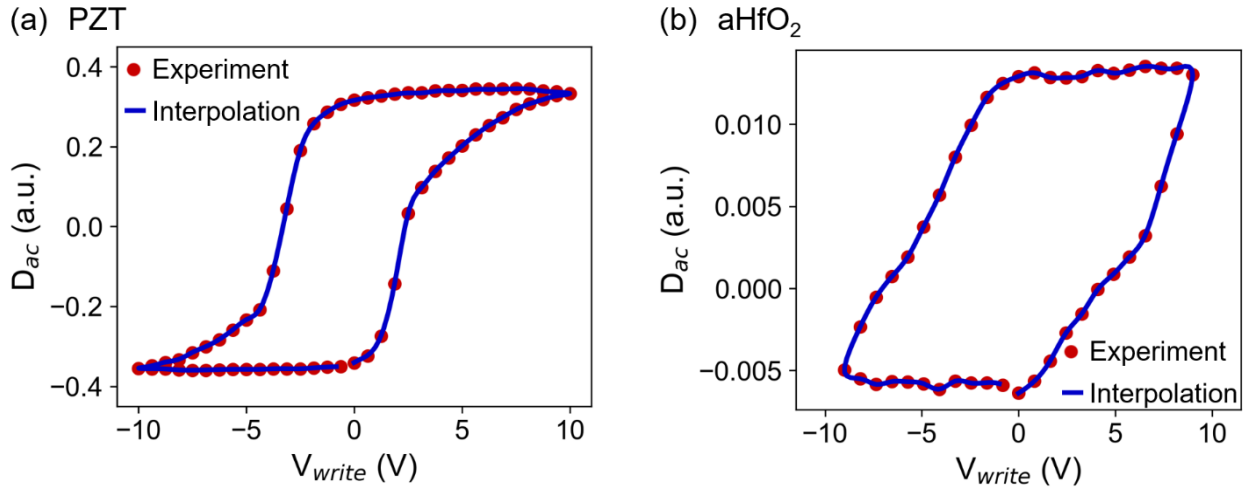

**Figure S1:** Experimental and interpolated response data measured (a) on PZT as a function of V$_{write}$ and (b) on aHfO$_2$. Data were extracted from the cKPFM maps shown in Figure 3 at V$_{read}$ = 0 V.

**Section II**

Maps of experimental and interpolated cKPFM data and their x- and y- gradients shown for different Vread and Vwrite amplitudes for PZT (Figure S2) and aHfO$_2$ (Figure S3).

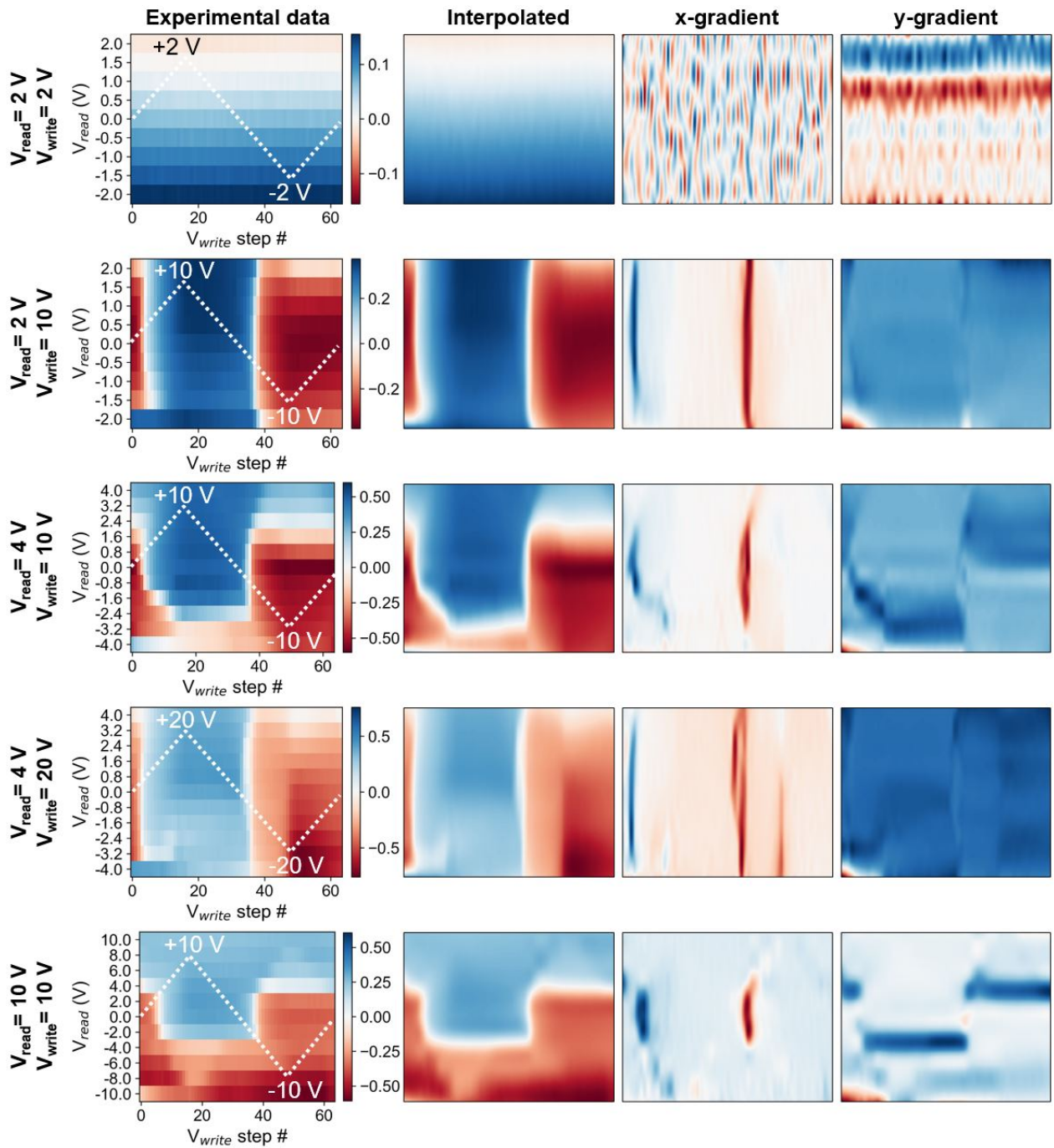

**Figure S2**: Maps of experimental cKPFM data acquired on PZT, interpolated data and the gradients in x- and y-direction calculated from interpolated maps. The $V_{read}$ and $V_{write}$ ratios vary according to the labels on the left. The first row shows response to sub-coercive read and write voltages.

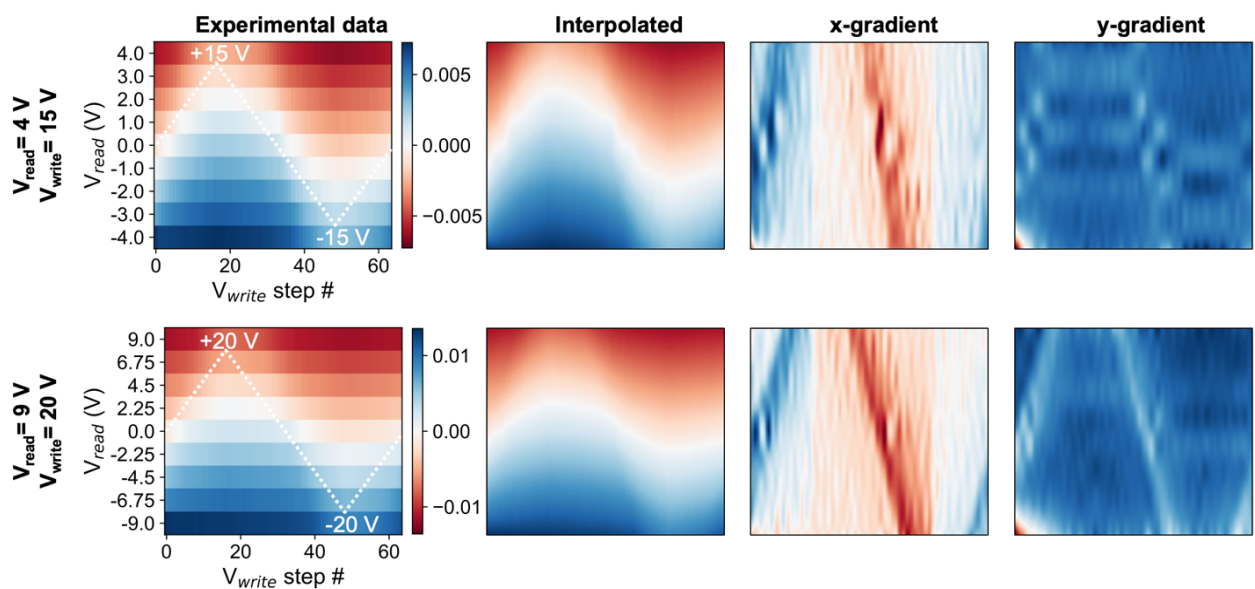

**Figure S3:** Maps of experimental cKPFM data acquired on aHfO$_2$, interpolated data and the gradients in x- and y-direction calculated from interpolated maps. The V$_{read}$ and V$_{write}$ amplitudes vary according to the labels on the left.

**Section III**

A multilayer perceptron ANN was trained on cKPFM maps within the data set shown in Figure 5(a), where 100 maps were acquired at 7 different temperature steps on PLZT across the ferroelectric – relaxor phase transition. Subsequently, the ANN was used to identify the temperature at which test data sets were measured.

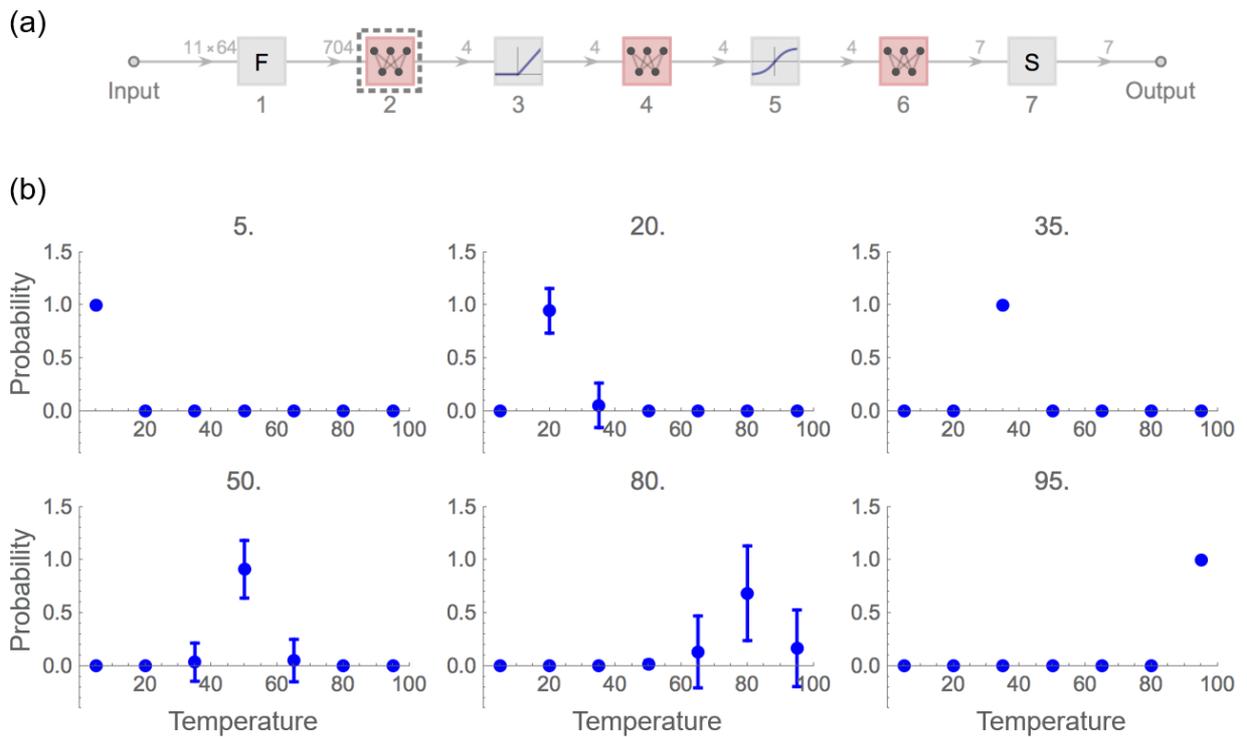

**Figure S4**: (a) Graph of linear multi-perceptron neural network used to predict temperatures based on the structure of the 2D response function. Layer structure of the network is as follows: (1) flattening image into a 704-dimensional vector; (2) mapping into 4 layer perceptron; (3) rectifying activation; (4) another mapping into 4 layer perceptron; (5) sigmoid activation; (6) mapping into 7 layer perceptron; (7) softmax normalization layer for normalization into class labels (each corresponding to specific temperature); (b) shows the average performance of the network to predict a specific temperature (labeled at the top) obtained by applying the network to validation data-set. (x – axis in each plot is the temperature in C; y-axis is the probability of predicted temperature)